\documentclass[aps,prl,amssymb,showpacs,floatfix,twocolumn,reprint,nofootinbib]{revtex4-1}
\usepackage{amsmath,amsfonts,graphics,epsfig,amssymb,color}

\begin{document}
\title{Unusual view of the Schwarzian theory }

\author{Vladimir V. Belokurov}

\email{vvbelokurov@yandex.ru}

\affiliation{Lomonosov Moscow State University, Leninskie gory 1, Moscow, 119991, Russia and Institute for Nuclear
Research of the Russian Academy of Sciences, 60th October Anniversary
Prospect 7a, Moscow, 117312, Russia}

\author{Evgeniy T. Shavgulidze}

\email{shavgulidze@bk.ru}

\affiliation{Lomonosov Moscow State University, Leninskie gory 1, Moscow, 119991, Russia}



\begin{center}
\begin{abstract}
A decomposition of the Wiener measure based on its quasi-invariance under the group of diffeomorphisms is proposed.
As a result, functional integrals in the Schwarzian theory can be written
as the Fourier transform of the integrals in a tahyonic model with the Calogero potential.
\end{abstract}
\end{center}
\maketitle

\vspace{1cm}

The Schwarzian theory is behind various physical models including the SYK model and the two-dimensional dilaton gravity (see, e.g., \cite{(Kit1)}, \cite{(Kit2)}, \cite{(MS)},   \cite{(GR)}, \cite{(MNW)}, \cite{(SW)}, \cite{(KitSuh)}, \cite{(Mertens)}, \cite{(HoMTV)}, and references therein).

The action of the theory is
\begin{equation}
   \label{Act2}
   I=-\frac{1}{\sigma^{2}}\int \limits _{0}^{1}\,\left[ \mathcal{S}_{\varphi}(t)+2\pi^{2}\left(\varphi'(t)\right)^{2}\right]dt\,,
\end{equation}
where
\begin{equation}
   \label{Der}
\mathcal{S}_{\varphi}(t)=
\left(\frac{\varphi''(t)}{\varphi'(t)}\right)'
-\frac{1}{2}\left(\frac{\varphi''(t)}{\varphi'(t)}\right)^2
\end{equation}
is the Schwarzian derivative, and $\varphi(t)$ is a diffeomorphism of the interval $[0,\,1]\,.$

An extraordinary universality of the Schwarzian theory is a consequence of its rich symmetry structure. At the same time, due to the invariance under the group of diffeomorphisms the theory needs a special handling. In \cite{(Sh)}, a quasi-invariant measure on the group of diffeomorphisms  was constructed. It has the
form
\begin{equation}
   \label{Measure}
   \mu_{\sigma}(d\varphi)=\exp\left\{\frac{1}{\sigma^{2}}\int \limits _{0}^{1}\, \mathcal{S}_{\varphi}(t)\,dt  \right\}  d\varphi\,.
\end{equation}
Under the substitution
\begin{equation}
   \label{subst}
 \varphi(t)=\frac{\int \limits _{0}^{t}\,\exp\{\xi(\tau)\}d\tau}{\int \limits _{0}^{1}\,\exp\{\xi(\eta)\}d\eta }  \,,
\end{equation}
the measure $\mu_{\sigma}(d\varphi)$ on the group $ Diff^{1}_{+}([0,\,1])$ turns into the Wiener measure  $w_{\sigma}(d\xi)$ on $C([0,\, 1])\,.$
Using the quasi-invariance of the measure (\ref{Measure}),   we have evaluated functional integrals for the partition function and the correlation functions in the Schwarzian theory explicitly \cite{(BShExact)}, \cite{(BShCorrelation)}.

 There are other fruitful approaches that use the symmetry of the Schwarzian  theory to link it to another theory where the corresponding calculations are much simpler than in the original theory
\cite{(SW)}, \cite{(BAK)}, \cite{(BAK2)}, \cite{(MTV)}, \cite{(GR2)}.

In this note. we propose a new representation of the functional integrals in the Schwarzian theory as the Fourier transform of the integrals in a tahyonic model with the Calogero potential.
To illustrate the capability of the proposed representation , we re-derive  the Schwarzian partition function
\begin{equation}
   \label{PF}
\int \limits _{ Diff_{+}^{1}([0, 1])/SL(2,\textbf{R}) }\exp\left\{-I \right\}  d\varphi\,.
\end{equation}

Note that  integrals over $Diff^{1} ([0,1])$ turn into the integrals over $Diff^{1}(S^{1})$ as follows \cite{(BShCorrelation)}:
$$
\int\limits_{Diff^{1} ([0,1]) }\delta\left(\frac{\varphi'(1)}{\varphi'(0)}-1 \right)\,F(\varphi)\,\mu_{\sigma}(d\varphi)
$$
\begin{equation}
   \label{IntCirc}
 =\frac{1}{\sqrt{2\pi}\sigma}\int\limits_{Diff^{1} (S^{1}) }F(\varphi)\mu_{\sigma}(d\varphi)
\,.
\end{equation}
The details of "glueing the ends of the interval" are given in \cite{(BShCorrelation)}. Here, we only mention that the proof of (\ref{IntCirc}) is based on the relation between the Wiener measure on the space of functions with an arbitrary right end and the measure on the space of Brownian bridges
\begin{equation}
   \label{Brown}
w_{\sigma}(d\zeta)=w^{Brownian}_{\sigma}(d\xi)\frac{1}{\sqrt{2\pi}\sigma}\exp\left(-\frac{\eta^{2}}{2\sigma^{2}} \right)d\eta\,,
\end{equation}
$$
\zeta(t)=\xi(t)\,,\ \ 0<t<1\,;\ \ \ \ \ \ \ \zeta(1)=\eta\,.
$$

Consider the Wiener measure with the variance 1 on the space of continuous positive functions on the circle $S^{1}$ (or on the interval $[0,\,1]$)
\begin{equation}
   \label{Wiener}
 w_{1}(dx)=\exp\left\{ -\frac{1}{2}\int \limits _{0}^{1}\,\left(x'(t) \right)^{2}dt \right\}\ dx\,.
\end{equation}

The measure (\ref{Wiener}) is quasi-invariant under the following action of the group of diffeomorphisms $Diff^{3}_{+} (S^{1})$ on $C_{+}(S^{1})$ \cite{(Shepp)}, \cite{(Sh)}, \cite{(BSh)}:
\begin{equation}
   \label{fx}
x\mapsto fx\,,\ \ (fx)(t)=x\left(f^{-1}(t) \right)\frac{1}{\sqrt{\left(f^{-1}(t) \right)'}}\,,
\end{equation}
$$
x\in C_{+}(S^{1}))\,,\ \ \  f\in Diff^{3}_{+}(S^{1})\,.
$$

The integral
\begin{equation}
   \label{inv}
\frac{\sigma^{2}}{4}=\int \limits _{0}^{1}\,\frac{1}{x^{2}(t)}dt
\end{equation}
is invariant under (\ref{fx}).

Define $\varphi\in Diff^{1}_{+} (S^{1})$ by the equation
\begin{equation}
   \label{fi}
\varphi^{-1}(t)=\frac{4}{\sigma^{2}}\,\int \limits _{0}^{t}\,\frac{1}{x^{2}(\tau)}d\tau +\theta\,.
\end{equation}

Now, with
\begin{equation}
   \label{xrofi}
x(t)=\frac{2}{\sigma}\,\frac{1}{\sqrt{\left(\varphi^{-1}(t)\right)'}}\,,
\end{equation}
we have a one-to-one correspondence
$$
(0,\,+\infty)\times Diff^{1}_{+} (S^{1})\leftrightarrow C_{+}(S^{1})\times S^{1}\,,
$$
that is, $
(\sigma\,,\ \varphi)\leftrightarrow(x\,,\ \theta)\,, \ \ \ \theta=\varphi^{-1}(0) \,.
$

We  assume that $x(0)=x(1)\,, \ \ \varphi'(0)=\varphi'(1)\,.$

For a smooth $ x(t)\ (x\in C^{1}_{+}(S^{1}))$ and $\varphi\in Diff^{3}_{+}(S^{1})\,,$
it is an easy exercise (see, e.g., \cite{(BSh)}) to verify that
\begin{equation}
   \label{intxprim}
 -\frac{1}{2}\int \limits _{0}^{1}\,\left(x'(t) \right)^{2}dt =\frac{1}{\sigma^{2}}\int \limits _{0}^{1}\,\mathcal{S}_{\varphi}(\tau)d\tau
\end{equation}
on the layer (\ref{inv}).

Thus we get the following ("polar") decomposition of the Wiener measure
\begin{equation}
   \label{MeasureDecomp}
w_{1}(dx)=\mathcal{P}(\sigma)\,\exp\left\{ \frac{1}{\sigma^{2}}\int \limits _{0}^{1}\,\mathcal{S}_{\varphi}(\tau)d\tau \right\}\ d\varphi\, d\sigma^{2}\,.
\end{equation}

Note that using (\ref{subst}) we can rewrite it as
\begin{equation}
   \label{Polar}
w_{1}(dx)=\mathcal{P}(\sigma)\,w_{\sigma}(d\xi)\, d\sigma^{2}\,.
\end{equation}

Now the following equality of functional integrals is valid:
$$
\int \limits _{ C_{+}(S^{1}) }F(x)\,\delta \left(\frac{\sigma^{2}}{4}-\int \limits _{0}^{1}\,\frac{1}{x^{2}(t)}dt \right)w_{1}(dx)
$$
$$
=\int \limits _{-\infty}^{+\infty}\,\mathcal{P}(\rho)\,\delta \left(\rho^{2}- \sigma^{2}\right)
$$
$$
\times \int \limits _{  Diff_{+}^{1}(S^{1}) }F(x(\rho,\,\varphi))\exp\left\{ \frac{1}{\rho^{2}}\int \limits _{0}^{1}\,\mathcal{S}_{\varphi}(\tau)d\tau \right\} d\varphi\, d\rho^{2}
$$
\begin{equation}
   \label{PolarWiener}
=\mathcal{P}(\sigma) \int \limits _{  Diff_{+}^{1}(S^{1}) }F(x(\sigma,\,\varphi))\,\exp\left\{ \frac{1}{\sigma^{2}}\int \limits _{0}^{1}\,\mathcal{S}_{\varphi}(\tau)d\tau \right\}\ d\varphi\,.
\end{equation}

The factor $\mathcal{P}(\sigma)$ can be used to normalize the Wiener measure of any layer (\ref{inv})
\begin{equation}
   \label{NormWienerLayer}
 \int \limits _{  C_{+}(S^{1}) }\delta \left(\frac{\sigma^{2}}{4}-\int \limits _{0}^{1}\,\frac{1}{x^{2}(t)}dt \right)w_{1}(dx)=\mathcal{P}(\sigma)\,.
\end{equation}
In this case,
\begin{equation}
   \label{NormMu}
   \int \limits _{  Diff_{+}^{1}(S^{1}) }\exp\left\{ \frac{1}{\sigma^{2}}\int \limits _{0}^{1}\,\mathcal{S}_{\varphi}(\tau)d\tau \right\} d\varphi=1\,.
\end{equation}

 Later on we will see that precisely this normalization corresponds to the normalization of the partition function used in \cite{(SW)}, \cite{(BShExact)}, \cite{(BShCorrelation)}.

To apply (\ref{PolarWiener}) in the Schwarzian theory, also note that (\ref{xrofi}) leads to
\begin{equation}
   \label{intxsq}
 \int \limits _{0}^{1}\,x^{2}(t) dt =\frac{4}{\sigma^{2}}\int \limits _{0}^{1}\,\left(\varphi'(\tau)\right)^{2}d\tau\,.
\end{equation}

To find the Schwarzian partition function, we evaluate
the regularised ($0 < \alpha < \pi $) functional integral
$$
Z_{\alpha}(\sigma)
$$
\begin{equation}
   \label{RegPF}
=\int \limits _{ Diff_{+}^{1}([0, 1]) }\exp\left\{\frac{1}{\sigma^{2}}\int \limits _{0}^{1}\,\left[ \mathcal{S}_{\varphi}(t)+2\alpha^{2}\dot{\varphi}^{2}(t)\right]dt  \right\}  d\varphi
\end{equation}
first.

According to (\ref{IntCirc}), (\ref{PolarWiener}), and (\ref{intxsq}), $Z_{\alpha}(\sigma)$ equals to
$$
\frac{1}{\sqrt{2\pi}\sigma}\, \mathcal{P}^{-1}(\sigma) \int \limits _{  C_{+}([0, 1]) }\exp\left\{\frac{\alpha^{2}}{2} \int \limits _{0}^{1}\,x^{2}(t) dt\right\}
$$
\begin{equation}
   \label{WienerPF}
\times\delta \left(\frac{\sigma^{2}}{4}-\int \limits _{0}^{1}\,\frac{1}{x^{2}(t)}dt \right)w_{1}(dx)\,.
\end{equation}

Note that due to (\ref{NormWienerLayer}),
\begin{equation}
   \label{Zeroalfa}
Z_{\alpha =0}(\sigma)=\frac{1}{\sqrt{2\pi}\sigma}\,.
\end{equation}

Taking the Fourier transform of the $\delta-$function in (\ref{WienerPF}), we get
$$
Z_{\alpha}(\sigma)=\frac{1}{\sqrt{2\pi}\sigma}\, \mathcal{P}^{-1}(\sigma)\, \frac{1}{2\pi}\int \limits _{-\infty}^{+\infty}\,\int \limits _{  C_{+}(S^{1}) }
$$
$$
\times\exp\left\{\int \limits _{0}^{1}\left[-\frac{1}{2}\left(x'(t) \right)^{2}+\frac{\alpha^{2}}{2}x^{2}(t)
-i\lambda\frac{1}{x^{2}(t)}\right] dt\right\}
$$
\begin{equation}
   \label{FourierTah}
\times dx\,\exp\left\{i\lambda\frac{\sigma^{2}}{4} \right\}\,d\lambda \,.
\end{equation}

The integral over $ C_{+}(S^{1}) $ in (\ref{FourierTah}) is nothing less than the functional integral for the partition function in the theory of a quantum oscillator with the Calogero potential given by the action
\begin{equation}
   \label{Tah}
A=\int \limits _{0}^{1}\left[-\frac{1}{2}\left(x'(t) \right)^{2}-\frac{m^{2}}{2}x^{2}(t)
-g\frac{1}{x^{2}(t)}\right] dt
\end{equation}
with the imaginary mass $m=i\alpha$ and an imaginary coupling constant $g=i\lambda \,.$

The solution of the quantum problem for the action (\ref{Tah}) is well known (see, e.g., \cite{(Perelomov)}) with the energy levels
$$
E_{n}=m\left(1+\sqrt{2g+\frac{1}{4}} \right)+2n\,m\,,\ \ \ \ \ \ \ n=0,\,1,\,2,\,...\,.
$$

Therefore, the partition function for the action (\ref{Tah}) equals to
$$
\int \limits _{  C_{+}(S^{1}) }\,e^{-A}\,dx=\sum \limits _{0}^{\infty}\exp\left\{-E_{n} \right\}
$$
\begin{equation}
   \label{TahPF}
=\exp\left\{-m\sqrt{2g+\frac{1}{4}} \right\}\frac{1}{2\sinh m}\,.
\end{equation}

After the simple integration over $\lambda $ we get the following result for
$Z_{\alpha}(\sigma)\,:$
$$
 Z_{\alpha}(\sigma)=\frac{\alpha}{\sin \alpha}\,\exp\left\{\frac{2\alpha^{2}}{\sigma^{2}} \right\}\,Z_{\alpha=0}(\sigma)
$$
\begin{equation}
   \label{Zalfa}
=\frac{\alpha}{\sin \alpha}\,\frac{1}{\sqrt{2\pi} \sigma}\,\exp\left\{ \frac{2\alpha^{2}}{\sigma^{2}}\right\}\,.
\end{equation}

The regularized volume $V_{\alpha}(\sigma) $ of the group $SL(2,\textbf{R})$ is \cite{(BShExact)}, \cite{(BShCorrelation)}
\begin{equation}
   \label{V}
V_{\alpha}(\sigma)=
=\exp \left\{-\frac{2\left[\pi^{2}-\alpha^{2}\right]}{\sigma^{2}}\right\} \frac{\pi \sigma^{2}}{\left[\pi^{2}-\alpha^{2}\right]}\,.
\end{equation}

Thus, for the Schwarzian partition function defined as the limit
\begin{equation}
   \label{Ratio2}
  Z_{Schw}(\sigma) = \lim \limits_{\alpha\rightarrow \pi - 0}\ \frac{Z_{\alpha}(\sigma)}{V_{\alpha}(\sigma)} \,,
\end{equation}
we get the known form:
\begin{equation}
   \label{Result2}
   Z_{Schw}(\sigma)= \frac{\sqrt{2\pi}}{\sigma^{3}}\,\exp\left\{ \frac{2\pi ^{2}}{\sigma^{2}}\right\}\,.
\end{equation}

The proposed decomposition of the Wiener measure (\ref{MeasureDecomp}) can be used not only for calculations in the Schwarzian theory. In the form (\ref{Polar}), it relates the Wiener measures with different variances, and represents the renormalization of the dynamical variable
$$
x(t)\rightarrow\frac{1}{\kappa}x(t)
$$
as a transfer to the corresponding layer.

\end{document}